\documentclass[12pt]{article}
\usepackage{graphicx,amssymb,amsmath,hyperref}

\newcommand{\be}{\begin{equation}}
\newcommand{\ee}{\end{equation}}
\newcommand{\bea}{\begin{eqnarray}}
\newcommand{\eea}{\end{eqnarray}}
\newcommand{\beas}{\begin{eqnarray*}}
\newcommand{\eeas}{\end{eqnarray*}}

\begin{document}
\begin{titlepage}

\vspace*{-24mm}

\rightline{NSF-KITP-12-105}

\vspace{4mm}

\begin{center}

{\Large Cosmic string interactions induced by gauge and scalar fields}

\vspace{4mm}

\renewcommand\thefootnote{\mbox{$\fnsymbol{footnote}$}}
Daniel Kabat${}^{1,2}$\footnote{daniel.kabat@lehman.cuny.edu} and
Debajyoti Sarkar${}^{1,3}$\footnote{dsarkar@gc.cuny.edu}

\vspace{4mm}

${}^1${\small \sl Department of Physics and Astronomy} \\
{\small \sl Lehman College of the CUNY, Bronx NY 10468, USA}

${}^2${\small \sl Kavli Institute for Theoretical Physics} \\
{\small \sl University of California, Santa Barbara CA 93106, USA}

${}^3${\small \sl Graduate School and University Center} \\
{\small \sl City University of New York, New York NY 10036, USA}

\end{center}

\vspace{4mm}

\noindent
We study the interaction between two parallel cosmic strings induced
by gauge fields and by scalar fields with non-minimal couplings to
curvature.  For small deficit angles the gauge field behaves like a
collection of non-minimal scalars with a specific value for the
non-minimal coupling.  We check this equivalence by computing the
interaction energy between strings at first order in the deficit
angles.  This result provides another physical context for the
``contact terms'' which play an important role in the renormalization
of black hole entropy due to a spin-1 field.

\end{titlepage}
\setcounter{footnote}{0}
\renewcommand\thefootnote{\mbox{\arabic{footnote}}}

\section{Introduction}

For a single cosmic string in four Euclidean dimensions the metric is \cite{Vilenkin:1981zs,Gott:1984ef}
\be
\label{cone}
ds^2 = dr^2 + r^2 d\psi^2 + d\tau^2 + dz^2
\ee
The string tension produces a deficit angle, $\psi \approx \psi + \beta$ where
\be
\beta = 2 \pi - 8 \pi \lambda
\ee
Here $\lambda = G \mu$ where $G$ is Newton's constant and $\mu$ is the mass per unit length of the string.

We will be interested in the interaction between two parallel cosmic
strings.  At the classical level there is no force between strings,\footnote{In classical gravity there is, however,
a non-trivial scattering amplitude which results from the conical boundary conditions \cite{'tHooft:1988yr,Deser:1988qn}.} but (as in the
Casimir effect) an interaction potential can be generated at one loop
by a quantum field propagating on this background.  For simplicity we
will take a perturbative approach, and calculate the interaction
energy at first order in the product of the two deficit angles.  We
consider two types of fields -- scalar fields with a non-minimal
coupling to curvature, and abelian gauge fields -- as the main point of this paper is to
highlight a relation between these two cases.
Vacuum polarization in the presence of a single cosmic string has been studied before; see for example \cite{Helliwell:1986hs,Linet:1987vz,Allen:1990mm} for scalar fields and \cite{Frolov:1987dz,Dowker:1987pk} for gauge fields.  For related calculations in the presence of multiple cosmic strings see \cite{Galtsov:1995xp,Aliev:1996va}.

We begin by recalling the argument that, to first order in the
background curvature, there should be a relation between gauge fields
and scalar fields with specific non-minimal couplings to curvature.
To our knowledge this relation was first stated in \cite{Larsen:1995ax},
although the essence of the following argument is taken from
\cite{Susskind:1994sm}.  Consider a spacetime which is a product
${\cal M}_n \times {\mathbb R}^{d-n}$ of a weakly-curved
$n$-dimensional Einstein manifold ${\cal M}_n$ with flat space
${\mathbb R}^{d-n}$.  The metric takes the form
\be
ds^2 = g_{\alpha\beta} dx^\alpha dx^\beta + \delta_{ij} dx^i dx^j
\ee
where $x^\alpha$ are coordinates on ${\cal M}_n$ and $x^i$ are
coordinates on ${\mathbb R}^{d-n}$.  The Einstein manifold has Ricci
curvature $R_{\alpha\beta} = {1 \over n} g_{\alpha\beta}
R$.\footnote{By Einstein manifold we mean a manifold with Ricci curvature locally
proportional to the metric, $R_{\alpha\beta}(x) = f(x)
g_{\alpha\beta}(x)$.  In two dimensions all manifolds are Einstein.  In
higher dimensions the contracted Bianchi identity $\nabla^\mu\big(R_{\mu\nu} -
{1 \over 2} g_{\mu\nu} R\big) = 0$ requires that $f$ be a constant.  In either case
it follows from the definition that $R_{\alpha\beta} = {1 \over n} g_{\alpha\beta} R$.}
Choose a vielbein $g_{\alpha \beta} = e_\alpha^a e_\beta^b
\delta_{ab}$ and denote the corresponding spin connection
$\omega_\alpha$.

To establish the relation between gauge and scalar fields we compare
their equations of motion.  For a gauge field, the equations of motion
in Feynman gauge are
\be
- \nabla_\nu \nabla^\nu A_\mu + R_{\mu \nu} A^\nu = 0
\ee
where $x^\mu = (x^\alpha,x^i)$.  There are ghosts associated with this
choice of gauge which behave like a pair of minimally-coupled scalar
fields \cite{Vassilevich:1994cz}.  The components of the gauge field
tangent to ${\mathbb R}^{d-n}$ obey
\be
- \nabla_\beta \nabla^\beta A_i - \partial_j \partial^j A_i = 0
\ee
where the covariant derivative $\nabla_\alpha$ treats $A_i$ as a
singlet of $SO(n)$.  That is, the components $A_i$ behave like
minimally-coupled scalar fields.  The components of the gauge field
tangent to ${\cal M}_n$, on the other hand, obey
\be
\label{AaEOM}
- \nabla_\beta \nabla^\beta A_a - \partial_j \partial^j A_a + {1 \over n} R A_a = 0
\ee
Here $\nabla_\alpha$ acts on $A_a = e_a{}^\alpha A_\alpha$ in the fundamental representation
of $SO(n)$, and we've made use of the fact that $R_{\alpha\beta} =
{1 \over n} g_{\alpha\beta} R$.  So the components $A_a$ are in the
fundamental representation of $SO(n)$ and have an explicit non-minimal
coupling to curvature.

Physical quantities can be computed perturbatively, as an expansion in
powers of the background curvature.  As a concrete example imagine
computing the effective action for the background which results from
integrating out $A_\mu$.  The spin connection can appear in the
effective action, but only through its field strength $F = d\omega +
\omega^2$.  In fact the field strength can first appear in the
effective action in terms such as $F_{\alpha\beta} F^{\alpha\beta}$
that are quadratic in the curvature.  So to first order in the
background curvature we can forget about the spin connection and treat
$A_a$ as a collection of $n$ scalar fields with a non-minimal coupling
to curvature.  The equation of motion for a non-minimal scalar is
\be
- \nabla_\beta \nabla^\beta \phi - \partial_j \partial^j \phi + \xi R \phi = 0\,,
\ee
and comparing to (\ref{AaEOM}) we identify the effective non-minimal
coupling parameter $\xi = 1/n$.  Thus to first order in the background
curvature a gauge field is equivalent to $n$ scalar fields with $\xi =
1/n$, plus $d-n$ minimally-coupled scalars.

This discussion is relevant to parallel cosmic strings because in two
dimensions every manifold is an Einstein manifold.  The argument
suggests that, to first order in the product of the deficit angles,
the interaction between two cosmic strings induced by a gauge field
should be the same as the interaction induced by an appropriate
collection of non-minimal scalars.

In the remainder this paper we verify this claim, by computing the
interaction energy between cosmic strings perturbatively.  In section
\ref{sect:scalar} we compute the interaction energy for a scalar
field, and in section \ref{sect:gauge} we carry out the corresponding
computation for a gauge field.  We conclude in section
\ref{sect:conclusions}, where we comment on our results and point out
the relation to studies of black hole entropy.

\section{Non-minimal scalar energy\label{sect:scalar}}

\begin{figure}
\begin{center}
\includegraphics{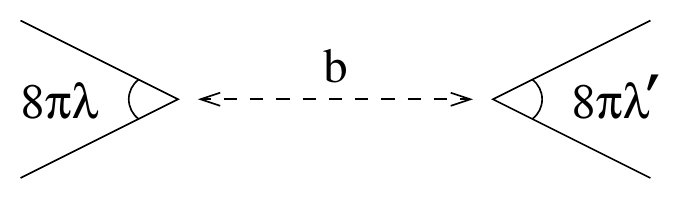}
\end{center}
\caption{Two parallel cosmic strings, separated by a distance $b$.\label{TwoStrings}}
\end{figure}

The Euclidean action is
\[
S = \int d^4x \, \sqrt{g} \left( {1 \over 2} g^{\mu\nu} \partial_\mu \phi \partial_\nu \phi + {1 \over 2} \xi R \phi^2 \right)
\]
For the conical geometry (\ref{cone}) the scalar curvature is\footnote{The easiest way to see
this is to note that a truncated cone, i.e.\ a disc with a conical
singularity at the center, has Euler characteristic $\chi = {1 \over
4 \pi} \int d^2x \sqrt{g} R + {1 \over 2 \pi} \beta = 1$.}
\be
\label{ScalarCurvature}
R = 16 \pi \lambda \delta^2(x) / \sqrt{g}
\ee
The action on a cone can be split into three pieces,
\be
\label{ScalarDecomposition}
S_{\rm cone} = S_0 + S_{\rm int}, \qquad S_{\rm int} = S_{\rm wedge} + S_{\rm tip}
\ee
where
\be
\label{free}
S_0 = \int d^4x \, {1 \over 2} \delta^{\mu\nu} \partial_\mu \phi \partial_\nu \phi
\ee
is the action in flat space,
\be
S_{\rm wedge} = - \int d\tau dz \, \int_0^\infty r dr\, \int_{-4\pi\lambda}^{4\pi\lambda} d\psi \,
{1 \over 2} \delta^{\mu\nu} \partial_\mu \phi \partial_\nu \phi
\ee
cancels the flat-space action in the region corresponding to the deficit angle, and
\be
S_{\rm tip} = \int d\tau dz \, 8 \pi \lambda \xi \phi^2
\ee
arises from the non-minimal coupling to curvature.  It's
straightforward to extend this to a pair of cosmic strings, just by
putting the deficit angles in opposite directions as shown in Fig.\
\ref{TwoStrings}.

We will treat $S_{\rm int}$ as a perturbation.\footnote{This is
somewhat subtle, since it's not manifest that perturbation theory in
$S_{\rm int}$ will enforce the proper conical boundary condition
$\phi(r,\psi) = \phi(r,\psi + \beta)$.  Fortunately the boundary
conditions are controlled by the spin connection on the cone, which
as we argued in the introduction can only enter at second order in
the deficit angle.}  To find the interaction energy per unit length
along the strings ${\cal H}_{\rm int}$ we use \cite{Ma}
\be
\int d\tau dz \, {\cal H}_{\rm int} = \langle 1 - e^{-S_{\rm int}} \rangle_{C,0}
\ee
where the subscript $C,0$ denotes a connected correlation function computed in the unperturbed theory (\ref{free}).
Expanding in powers of $S_{\rm int}$, the leading ${\cal O}(\lambda \lambda')$ interaction between the strings comes from
\be
\int d\tau dz \, {\cal H}_{\rm int} \approx - \langle S_{\rm int}^{(1)} \, S_{\rm int}^{(2)} \rangle_{C,0}
\ee
where the superscripts (1), (2) refer to the first and second cosmic string, respectively.  Some useful unperturbed correlators are
\be
\langle \phi(x) \phi(x') \rangle = {1 \over 4 \pi^2} \, {1 \over (x - x')^2}
\ee
and
\beas
&& \big\langle (\partial \phi)^2 (x) \, (\partial \phi)^2 (x') \big\rangle = {6 \over \pi^4} \, {1 \over (x - x')^8} \\
&& \big\langle (\partial \phi)^2 (x) \, \phi^2 (x') \big\rangle = {1 \over 2 \pi^4} \, {1 \over (x - x')^6} \\
&& \big\langle \phi^2(x) \, \phi^2(x') \big\rangle = {1 \over 8 \pi^4} \, {1 \over (x - x')^4}
\eeas

There are three types of interactions.  For generality we can imagine that the
two strings have different non-minimal couplings $\xi$, $\xi'$.

\goodbreak
\noindent
\underline{wedge -- wedge}

To first order in $\lambda$ and $\lambda'$ the wedges can be treated as very narrow, so that
\beas
{\cal H}_{\rm int} & = & - 16 \pi^2 \lambda \lambda' \int d\tau dz \, \int_0^\infty x dx \int_0^\infty x' dx' \,
{6 \over \pi^4} \, {1 \over \big(\tau^2 + z^2 + (x + x' + b)^2\big)^4} \\
& = & - {4 \lambda \lambda' \over 15 \pi b^2}
\eeas

\goodbreak
\noindent
\underline{wedge -- tip}

For wedge 1 with tip 2 we have
\beas
{\cal H}_{\rm int} & = & 32 \pi^2 \lambda \lambda' \xi' \int d\tau dz \, \int_0^\infty x dx \,
{1 \over 2 \pi^4} \, {1 \over \big(\tau^2 + z^2 + (x + b)^2\big)^3} \\
& = & {4 \lambda \lambda' \xi' \over 3 \pi b^2}
\eeas

\goodbreak
\noindent
\underline{tip -- tip}

The interaction between the two tips is
\beas
{\cal H}_{\rm int} & = & - 64 \pi^2 \lambda \lambda' \xi \xi' \int d\tau dz \,
{1 \over 8 \pi^4} \, {1 \over \big(\tau^2 + z^2 + b^2\big)^2} \\
& = & - {8 \lambda \lambda' \xi \xi' \over \pi b^2}
\eeas

\noindent
Assembling these results, to first order in $\lambda$ and $\lambda'$ the interaction energy per unit length due to
a non-minimally coupled scalar field is
\be
\label{Escalar}
{\cal H}_{\rm int} = {\lambda \lambda' \over \pi b^2} \left(- {4 \over 15} + {4 \over 3}(\xi + \xi') - 8 \xi \xi'\right)
\ee

To check the validity of our perturbative approach consider computing
$\langle \phi^2 \rangle$ for a minimally-coupled scalar field in the
presence of a single cosmic string.  At first order in perturbation
theory, after subtracting the divergence which is present in flat
space, we have
\be
\label{phi2}
\langle \phi^2 \rangle = - \langle \phi^2 S_{\rm wedge} \rangle_{C,0} = {\lambda \over 6 \pi^2 r^2}
\ee
where $r$ is the distance from the tip of the cone.  On the other hand $\langle \phi^2 \rangle$ can be computed exactly,
\be
\langle \phi^2 \rangle = \int_0^\infty ds \, K(s,x,x)
\ee
where the scalar heat kernel on a cone is\footnote{See for example
\cite{Kabat:1995eq}.  We dropped the term in the heat kernel $1 /
(4 \pi s)^2$ which is responsible for the divergence in flat space.}
\be
K(s,x,x) = - {1 \over 2 \beta} {1 \over (4 \pi s)^2} \int_{-\infty}^\infty dy \, e^{-{r^2\over s} \cosh^2(y/2)}
\left(\cot {\pi \over \beta} (\pi + i y) + \cot {\pi \over \beta} (\pi - i y)\right)
\ee
Expanding the heat kernel to first order in the deficit angle and integrating over $s$ reproduces (\ref{phi2}).

\section{Gauge field energy\label{sect:gauge}}

We start from the Euclidean action
\beas
S & = & S_{\hbox{\small\rm Maxwell}} + S_{\hbox{\small\rm gauge fixing}} \\
& = & \int d^dx \sqrt{g} \left( {1 \over 4} F_{\mu\nu} F^{\mu\nu} + {1 \over 2} \big(\nabla_\mu A^\mu\big)^2 \right)
\eeas
There are ghosts associated with this choice of gauge that behave like a pair of minimally-coupled scalars.

If we smooth out the conical singularities, so that we can freely integrate by parts, the action becomes
\beas
S & = & \int d^dx \sqrt{g} \left({1 \over 2} \nabla_\mu A_\nu \nabla^\mu A^\nu
- {1 \over 2} A^\mu \big(\nabla_\mu \nabla_\nu - \nabla_\nu \nabla_\mu\big) A^\nu\right) \\
& = & \int d^dx \sqrt{g} \left({1 \over 2} \nabla_\mu A_\nu \nabla^\mu A^\nu + {1 \over 2} R_{\mu\nu} A^\mu A^\nu \right)
\eeas
In the second line we used $[\nabla_\mu, \nabla_\nu] A^\nu = -
R_{\mu\nu} A^\nu$.  We work on a space which is a product of a
two-dimensional cone with coordinates $x^\alpha$ and a
$(d-2)$-dimensional flat space with coordinates $x^i$.
\[
ds^2 = g_{\alpha\beta} dx^\alpha dx^\beta + \delta_{ij} dx^i dx^j
\]
In two dimensions the Ricci tensor is proportional to the metric, so from (\ref{ScalarCurvature})
\be
R_{\alpha\beta} = 8 \pi \lambda g_{\alpha\beta} \delta^2(x) / \sqrt{g}
\ee
where $8 \pi \lambda$ is the deficit angle.  Thus the action for a gauge field on a cone can be decomposed into
\be
S_{\rm cone} = S_0 + S_{\rm int}, \qquad S_{\rm int} = S_{\rm wedge} + S_{\rm tip}
\ee
For example in four dimensions
\be
S_0 = \int d^4x \, {1 \over 2} (\partial_\mu A_\nu)^2
\ee
is the Feynman gauge action in flat space,
\be
S_{\rm wedge} = - \int d\tau dz \, \int_0^\infty r dr\, \int_{-4\pi\lambda}^{4\pi\lambda} d\psi \, {1 \over 2} (\partial_\mu A_\nu)^2
\ee
cancels the flat-space action in the region corresponding to the deficit angle, and
\be
S_{\rm tip} = 4 \pi \lambda \int d\tau dz \, g_{\alpha\beta} A^\alpha A^\beta
\ee
arises from the explicit coupling to curvature.  Aside from the sums over photon polarizations, this is identical to the decomposition of the non-minimal
scalar action (\ref{ScalarDecomposition}).

The interaction between two cosmic strings can be calculated
perturbatively, just as for a non-minimal scalar field.\footnote{Again
it's not manifest that perturbation theory in $S_{\rm int}$ enforces
the proper conical boundary conditions on $A_\alpha$, but this
effect is controlled by the spin connection which can only enter at
second order in the deficit angle.} In fact the two calculations are
identical.  There are $d-2$ polarizations transverse to the cone which
behave in perturbation theory just like minimally-coupled scalars.  Two of these
polarizations are canceled by the ghosts, leaving no contribution in
four dimensions.  The two polarizations tangent to the cone behave
like non-minimal scalars with $\xi = 1/2$.  So the overall interaction
energy coming from a gauge field in four dimensions is simply twice
the scalar result (\ref{Escalar}) evaluated at $\xi = 1/2$.  That is,
for a gauge field in four dimensions
\be
{\cal H}_{\rm int} = {2 \lambda \lambda' \over \pi b^2} \left(- {14 \over 15}\right)
\ee

To check the validity of our perturbative approach consider computing $\langle
A_\mu A^\mu \rangle$ around a single cosmic string.  In perturbation
theory, after subtracting the divergence present in flat space, we
have
\be
\label{A2}
\langle A_\mu A^\mu \rangle = \langle A_\mu A^\mu \left(-S_{\rm wedge} - S_{\rm tip}\right) \rangle_{C,0}
= {4 \lambda \over 6 \pi^2 r^2} - {\lambda \over \pi^2 r^2}
\ee
The first term comes from $S_{\rm wedge}$ and is four times the scalar
field result (\ref{phi2}).  The second term comes from $S_{\rm tip}$
and reflects the non-minimal coupling to curvature.  The same quantity
can be computed exactly,
\be
\langle A_\mu A^\mu \rangle = \int_0^\infty ds \, g_{\mu\nu} K^{\mu\nu}_{\rm vector}(s,x,x)
\ee
where the vector heat kernel is \cite{Kabat:1995eq}
\be
\label{Kvector}
g_{\mu\nu} K^{\mu\nu}_{\rm vector} = 4 K_{\rm scalar}(s,x,x) + {2 \over r} \partial_r s K_{\rm scalar}(s,x,x)
\ee
Expanding to first order in the deficit angle and integrating over $s$
reproduces (\ref{A2}).\footnote{Note that the last term in
(\ref{Kvector}), which in the black hole context captures the
contact interaction of a gauge field with the horizon, corresponds
at first order in perturbation theory to effects associated with
$S_{\rm tip}$.}

\section{Conclusions\label{sect:conclusions}}

In this paper we considered a cosmic string spacetime and argued that
to first order in the deficit angle there is an equivalence between a
gauge field and a collection scalar fields with specific non-minimal
couplings to curvature.  More generally the equivalence holds on the
product of any weakly-curved Einstein manifold with flat space.  We
tested the equivalence by computing the interaction energy between two
cosmic strings to first order in perturbation theory, showing that it
indeed matched for the appropriate value of the non-minimal coupling
parameter.

Throughout this paper we worked in Feynman gauge, which is adequate
for studying gauge-invariant quantities.  However it would be
interesting to study the relation between gauge and scalar fields in
other choices of gauge.  Also it would be interesting to study the
interaction between strings at higher orders in perturbation theory.
Beyond leading order there is no reason to expect an equivalence
between gauge and scalar fields, since the spin connection
distinguishes between the two types of fields and can appear in the
interaction energy at second order in the deficit angle.

Besides their direct application to cosmic strings, our results also
have relevance to the thermodynamics of black holes.  In a Euclidean
formalism the entropy of a black hole measures the response of the
partition function to an infinitesimal conical deficit angle inserted
at the horizon \cite{Susskind:1993ws,Carlip:1993sa}.  This has been
used to study the renormalization of black hole entropy due to matter
fields, with the somewhat surprising conclusion that a gauge field can
make a negative contribution to the entropy.  In \cite{Kabat:1995eq}
it was shown that this is due to a contact term in the partition
function for a gauge field, associated with particle paths that begin
and end on the horizon.  Here we've shown that, to first order in the
deficit angle, a gauge field is equivalent to a collection of
non-minimal scalars.  So the contact interaction of
\cite{Kabat:1995eq} is visible at the level of the equations of
motion, as the explicit non-minimal coupling to curvature seen in
(\ref{AaEOM}).  This makes the negative renormalization of black hole
entropy less mysterious, since it maps a gauge field to the
well-studied problem of a non-minimally coupled scalar field in a
black hole background \cite{Solodukhin:1995ak}.  Our results also show
the physical relevance of these contact interactions: besides
contributing to black hole entropy, they make a (finite, observable,
gauge invariant) contribution to the force between two cosmic strings.

We conclude with some additional evidence in support of the relation
between gauge and scalar fields at first order in the background
curvature.  The partition function for a gauge field on a cone was
evaluated in \cite{Kabat:1995eq}.  Including the ghosts, the result is
\be
\beta F_{\rm gauge} = (d-2) \beta F^{\rm minimal}_{\rm scalar} + A_\perp (2 \pi - \beta)
\int_{\epsilon^2}^\infty {ds \over (4 \pi s)^{d/2}}
\ee
Here $d$ is the total number of spacetime dimensions, $A_\perp$ is the
area of the $d-2$ transverse dimensions corresponding to the horizon,
$s$ is a Schwinger parameter, and $\epsilon$ is a UV cutoff.  The
partition function for a non-minimal scalar was evaluated to first
order in the deficit angle in \cite{Solodukhin:1995ak}, with the
result
\be
\beta F^{\,\xi}_{\rm scalar} = \beta F^{\rm minimal}_{\rm scalar} + \xi A_\perp (2 \pi - \beta)
\int_{\epsilon^2}^\infty {ds \over (4 \pi s)^{d/2}}
\ee
Comparing the partition functions again shows that a gauge field
corresponds to two non-minimal scalars with $\xi = 1/2$, together with
$d-2$ minimal scalars (two of which are canceled by the ghosts).
The same relation can be seen in the one-loop renormalization of Newton's constant,
\be
{1 \over 4 G_{N,{\rm ren}}} = {1 \over 4 G_N} + {c_1 \over (4 \pi)^{d - 2 \over 2} (d - 2) \epsilon^{d-2}}
\ee
where the Seeley -- de Witt coefficients are \cite{Birrell:1982ix}
\be
c_1 = \left\lbrace\begin{array}{ll}
{1 \over 6} - \xi & \hbox{\rm non-minimal scalar} \\[5pt]
{d - 2 \over 6} - 1 & \hbox{\rm gauge field including ghosts}
\end{array}\right.
\ee
On a $d$-dimensional Einstein manifold the gauge field result
corresponds to $d$ non-minimal scalars with $\xi = 1/d$, plus two
minimally-coupled scalar ghosts.

\bigskip
\goodbreak
\centerline{\bf Acknowledgements}
\noindent
We are grateful to Dario Capasso, Ted Jacobson and Aron Wall for
valuable discussions.  This work was supported in part by U.S.\ National
Science Foundation grants PHY-0855582 and PHY11-25915 and by PSC-CUNY
grants.


\begin{thebibliography}{10}

\bibitem{Vilenkin:1981zs}
A.~Vilenkin, ``{Gravitational field of vacuum domain walls and strings},''
\href{http://dx.doi.org/10.1103/PhysRevD.23.852}{{\em Phys.Rev.} {\bfseries D23} (1981) 852--857}.

\bibitem{Gott:1984ef}
J.R.~Gott, ``{Gravitational lensing effects of vacuum strings: Exact solutions},''
\href{http://dx.doi.org/10.1086/162808}{{\em Astrophys.J.} {\bfseries 288} (1985) 422--427}.

\bibitem{'tHooft:1988yr}
G.~'t Hooft,
``Nonperturbative two particle scattering amplitudes in (2+1)-dimensional quantum gravity,''
Commun.\ Math.\ Phys.\  {\bf 117}, 685 (1988).
  
\bibitem{Deser:1988qn} 
S.~Deser and R.~Jackiw,
``Classical and quantum scattering on a cone,''
Commun.\ Math.\ Phys.\  {\bf 118}, 495 (1988).

\bibitem{Helliwell:1986hs} 
  T.~M.~Helliwell and D.~A.~Konkowski,
  ``Vacuum fluctuations outside cosmic strings,''
  Phys.\ Rev.\ D {\bf 34}, 1918 (1986).
  
\bibitem{Linet:1987vz} 
  B.~Linet,
  ``Quantum field theory in the space-time of a cosmic string,''
  Phys.\ Rev.\ D {\bf 35}, 536 (1987).
  
\bibitem{Allen:1990mm} 
  B.~Allen and A.~C.~Ottewill,
  ``Effects of curvature couplings for quantum fields on cosmic string space-times,''
  Phys.\ Rev.\ D {\bf 42}, 2669 (1990).
  
\bibitem{Frolov:1987dz} 
  V.~P.~Frolov and E.~M.~Serebryanyi,
  ``Vacuum polarization in the gravitational field of a cosmic string,''
  Phys.\ Rev.\ D {\bf 35}, 3779 (1987).
  
\bibitem{Dowker:1987pk} 
  J.~S.~Dowker,
  ``Vacuum averages for arbitrary spin around a cosmic string,''
  Phys.\ Rev.\ D {\bf 36}, 3742 (1987).
  
\bibitem{Galtsov:1995xp} 
  D.~V.~Galtsov, Y.~V.~Grats and A.~B.~Lavrentev,
  ``Vacuum polarization and topological selfinteraction of a charge in multiconic space,''
  Phys.\ Atom.\ Nucl.\  {\bf 58}, 516 (1995)
  [Yad.\ Fiz.\  {\bf 58}, 570 (1995)].
  
\bibitem{Aliev:1996va} 
  A.~N.~Aliev,
  ``Casimir effect in the space-time of multiple cosmic strings,''
  Phys.\ Rev.\ D {\bf 55}, 3903 (1997).

\bibitem{Larsen:1995ax}
F.~Larsen and F.~Wilczek, ``{Renormalization of black hole entropy and of the gravitational coupling constant},''
\href{http://dx.doi.org/10.1016/0550-3213(95)00548-X}{{\em Nucl.Phys.} {\bfseries B458} (1996) 249--266},
\href{http://arxiv.org/abs/hep-th/9506066}{{\ttfamily arXiv:hep-th/9506066 [hep-th]}}.

\bibitem{Susskind:1994sm}
L.~Susskind and J.~Uglum, ``{Black hole entropy in canonical quantum gravity and superstring theory},''
\href{http://dx.doi.org/10.1103/PhysRevD.50.2700}{{\em Phys.Rev.} {\bfseries D50} (1994) 2700--2711},
\href{http://arxiv.org/abs/hep-th/9401070}{{\ttfamily arXiv:hep-th/9401070 [hep-th]}}.

\bibitem{Vassilevich:1994cz}
See for example
D.~V. Vassilevich, ``{QED on curved background and on manifolds with boundaries: Unitarity versus covariance},''
\href{http://dx.doi.org/10.1103/PhysRevD.52.999}{{\em Phys.Rev.} {\bfseries D52} (1995) 999--1010},
\href{http://arxiv.org/abs/gr-qc/9411036}{{\ttfamily arXiv:gr-qc/9411036 [gr-qc]}}.

\bibitem{Ma}
S.-K.~Ma, {\em Statistical Mechanics} (World Scientific, 1985).  See page 223.

\bibitem{Kabat:1995eq}
D.~Kabat, ``{Black hole entropy and entropy of entanglement},''
\href{http://dx.doi.org/10.1016/0550-3213(95)00443-V}{{\em Nucl.Phys.} {\bfseries B453} (1995) 281--302},
\href{http://arxiv.org/abs/hep-th/9503016}{{\ttfamily arXiv:hep-th/9503016 [hep-th]}}.

\bibitem{Susskind:1993ws}
L.~Susskind, ``{Some speculations about black hole entropy in string theory},''
\href{http://arxiv.org/abs/hep-th/9309145}{{\ttfamily arXiv:hep-th/9309145 [hep-th]}}.

\bibitem{Carlip:1993sa}
S.~Carlip and C.~Teitelboim, ``{The off-shell black hole},''
\href{http://dx.doi.org/10.1088/0264-9381/12/7/011}{{\em Class.Quant.Grav.} {\bfseries 12} (1995) 1699--1704},
\href{http://arxiv.org/abs/gr-qc/9312002}{{\ttfamily arXiv:gr-qc/9312002 [gr-qc]}}.

\bibitem{Solodukhin:1995ak}
S.~N. Solodukhin, ``{One loop renormalization of black hole entropy due to nonminimally coupled matter},''
\href{http://dx.doi.org/10.1103/PhysRevD.52.7046}{{\em Phys.Rev.} {\bfseries D52} (1995) 7046--7052},
\href{http://arxiv.org/abs/hep-th/9504022}{{\ttfamily arXiv:hep-th/9504022 [hep-th]}}.

\bibitem{Birrell:1982ix}
N.~D.~Birrell and P.~C.~W.~Davies, {\em Quantum Fields In Curved Space} (Cambridge, 1982).  See section 6.2.

\end{thebibliography}

\providecommand{\href}[2]{#2}\begingroup\raggedright\endgroup

\end{document}